\documentclass[aps,superscriptaddress,twocolumn,showpacs,showkeys]{revtex4}
\usepackage{graphicx}
\usepackage{amsmath}

\begin{document}

\title{Matrix representation of evolving networks\footnote{presented at the First Polish 
Symposium on Econo- and Sociophysics, Warsaw, Nov. 19-20, 2004}}

\author{Krzysztof Malarz}
\homepage{http://home.agh.edu.pl/malarz/}

\author{Krzysztof Ku{\l}akowski}
\email{kulakowski@novell.ftj.agh.edu.pl}
\affiliation{
Faculty of Physics and Applied Computer Science,
AGH University of Science and Technology,\\
al. Mickiewicza 30, PL-30059 Krak\'ow, Poland.
}

\date{\today}

\begin{abstract}
We present the distance matrix evolution for different types of networks: 
exponential, scale-free and classical random ones. Statistical properties of 
these matrices are discussed as well as topological features of the networks. Numerical data
on the degree and distance distributions are compared with theoretical predictions.
\end{abstract}

\pacs{
02.10.Ox, 
05.10.-a, 
07.05.Tp  
}

\keywords{computer modeling and simulation; complex networks; trees and graphs; sociophysics; 
econophysics}

\maketitle

\section{Introduction}
The problem of evolving networks \cite{graph,a-b-rew,d-m-rew,newman-rew,d-m-s-rew} belongs 
to a~new area of statistical physics with many interdisciplinary applications, from biology 
(sexual contacts, food webs, ecological networks) \cite{bio} via sociophysics (the strength 
of weak ties, terrorism, scientific collaborations, paper citations networks) \cite{socio},
econophysics (agents' games and interactions, business contacts) \cite{econo},
to computer science (Internet infrastructure and world wide web) \cite{cmpsc}.

In all these cases a central role is played by {\em graphs} which allow to describe networks 
mathematically.
A graph is a set $\mathcal{V}$ of $N$ {\em vertices} and a set $\mathcal{E}$ of $L$ {\em edges} 
among vertices.
A simple graph is a graph without loops (i.e. self-links) and without multiple edges.
A forest is a graph without cycles (i.e. paths which start and end at the same node).
Connected forest is called a tree \cite{graph}.
The terminology depends on subject where graph theory is applied: vertices become nodes, 
actors and agents in computer science, socio- and econophysics, respectively.
Edges are called links or interactions as well \cite{a-b-rew,d-m-rew,newman-rew}.
By {\em evolving} we mean adding subsequent nodes to an already existing graph with $M$ 
links to $M$ preexisting nodes.
For $M=1$ a tree, and for $M=2$ a simple graph appears.
Here evolution means {\em growth}.

Other evolution strategy is to modify an existing network without its growth, i.e. without 
increasing number of elements of the set $\mathcal{V}$.
Formation of a network may take place via {\em the rewiring} procedure \cite{watts}: starting 
with a ring of $N$ connected nodes --- each of them having $K$ nearest neighbors --- we destroy 
a randomly selected link $i$---$m$ and create instead another link $i$---$n$.

Another strategy is to start with $N$ nodes and $L$ edges between them.
A structure formed in this way is termed Erd\H os--\-R\'enyi graph \cite{crg}.
Similar approach was proposed by Gilbert \cite{gilbert} where number $N$ of nodes is fixed and 
a new link between each of $N(N-1)/2$ pair is realized with given probability $p$.
For $N\to\infty$ Gilbert and Erd\H os--\-R\'enyi models give the same results and 
$p=2LN^{-1}(N-1)^{-1}$.
Graphs described above are called {\em classical random graphs} (CRG) \cite{gilbert}.
The graph ``{\em thermalization}'' is a generalization of the rewiring strategy accompanied 
by Metropolis dynamics \cite{burda-1}.

Let us denote $P_a(m)$ the probability that a new node will be attached to an existing node $m$.
For {\em the scale-free} (preferential, Albert--Barab\'asi) networks \cite{ab-org}, $P_a(m)$ is 
proportional to the node degree $k(m)$ (i.e. number of edges which leads 
from/to $m$) \cite{a-b-rew,d-m-rew,newman-rew}.
For {\em the exponential} networks, $P_a(m)$ is uniform \cite{a-b-rew,d-m-rew}.
 
\begin{figure*}
\begin{center}
(a) \includegraphics[width=.4\textwidth]{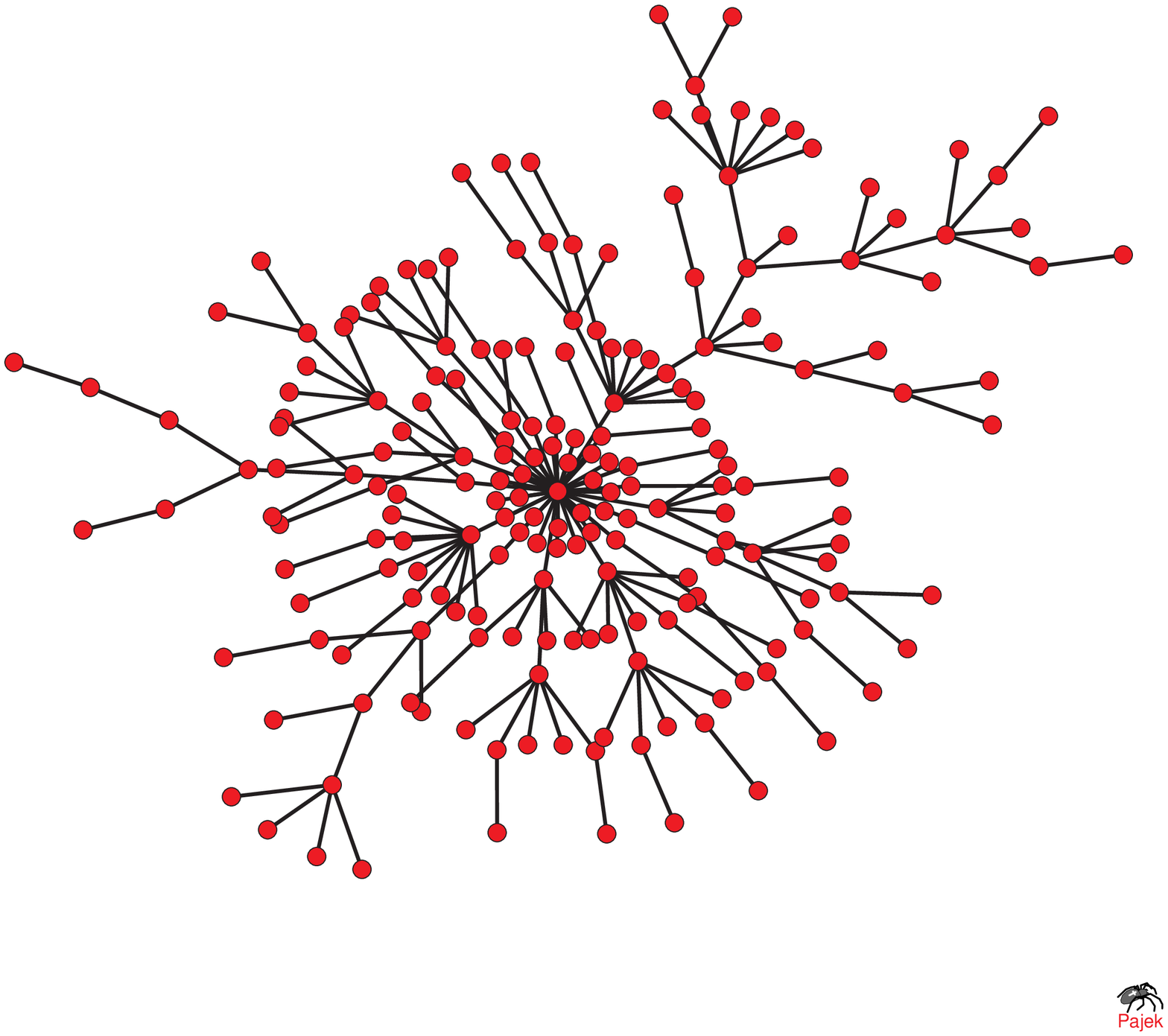}
(b) \includegraphics[width=.4\textwidth]{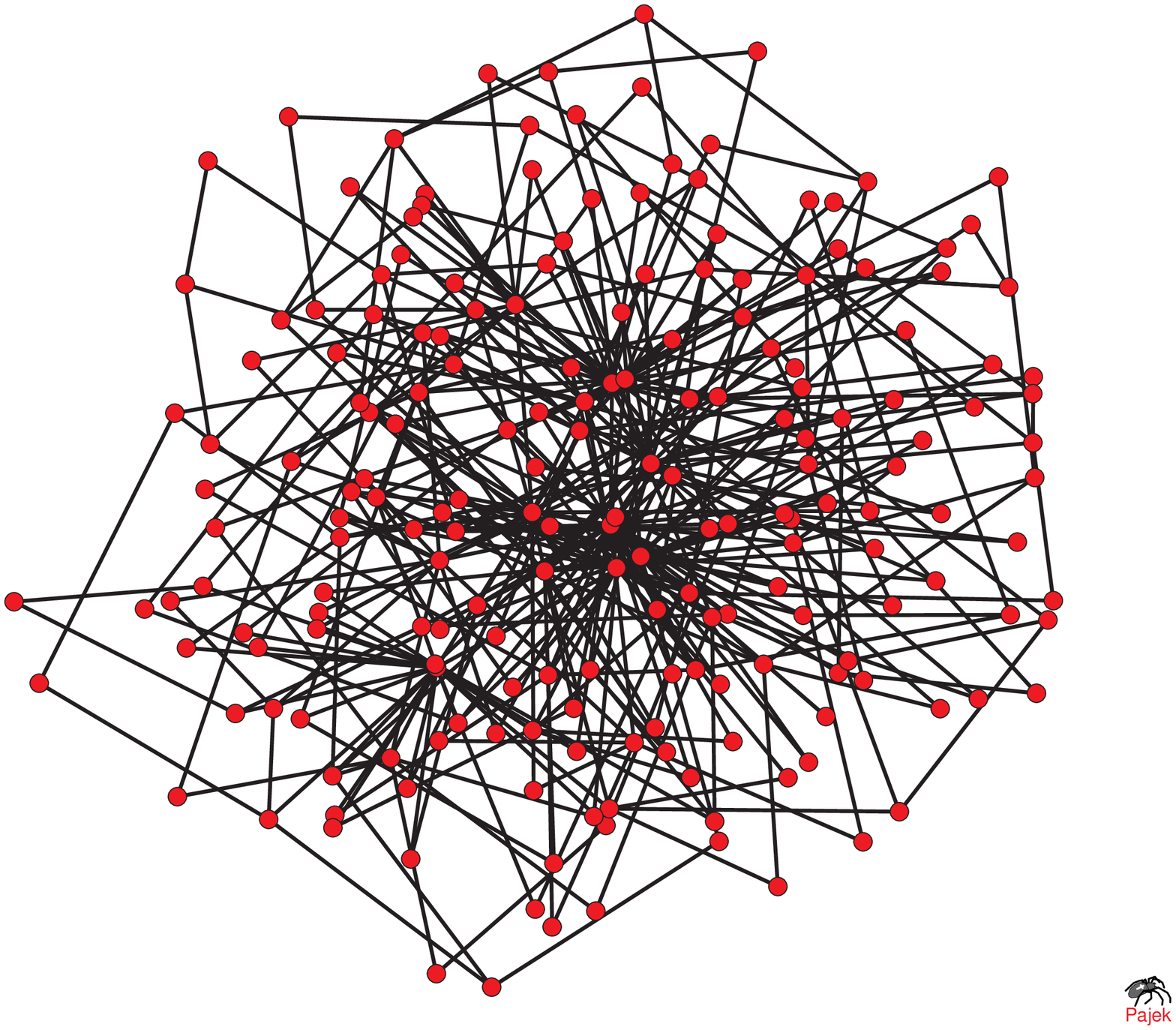}\\
(c) \includegraphics[width=.4\textwidth]{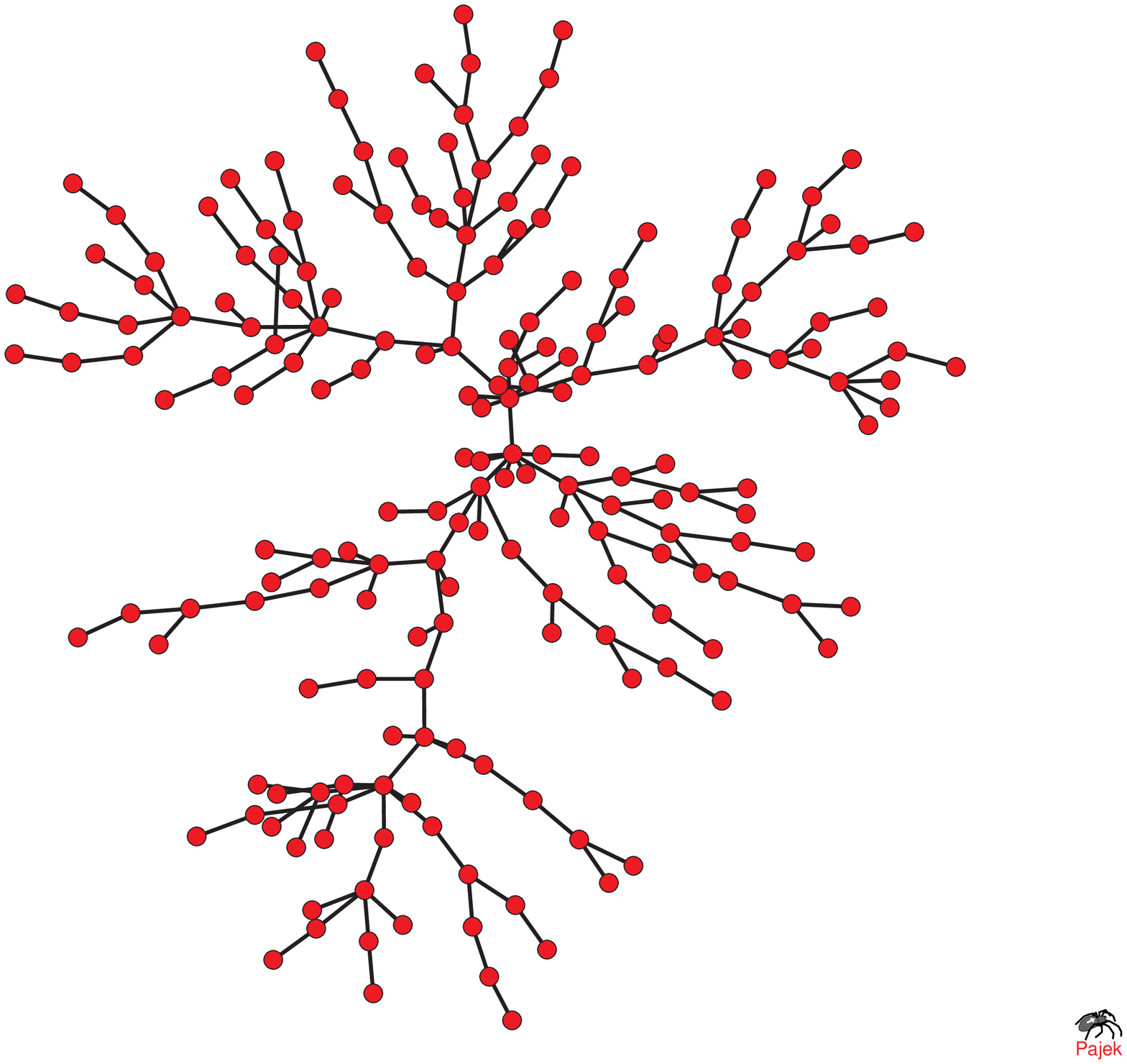}
(d) \includegraphics[width=.4\textwidth]{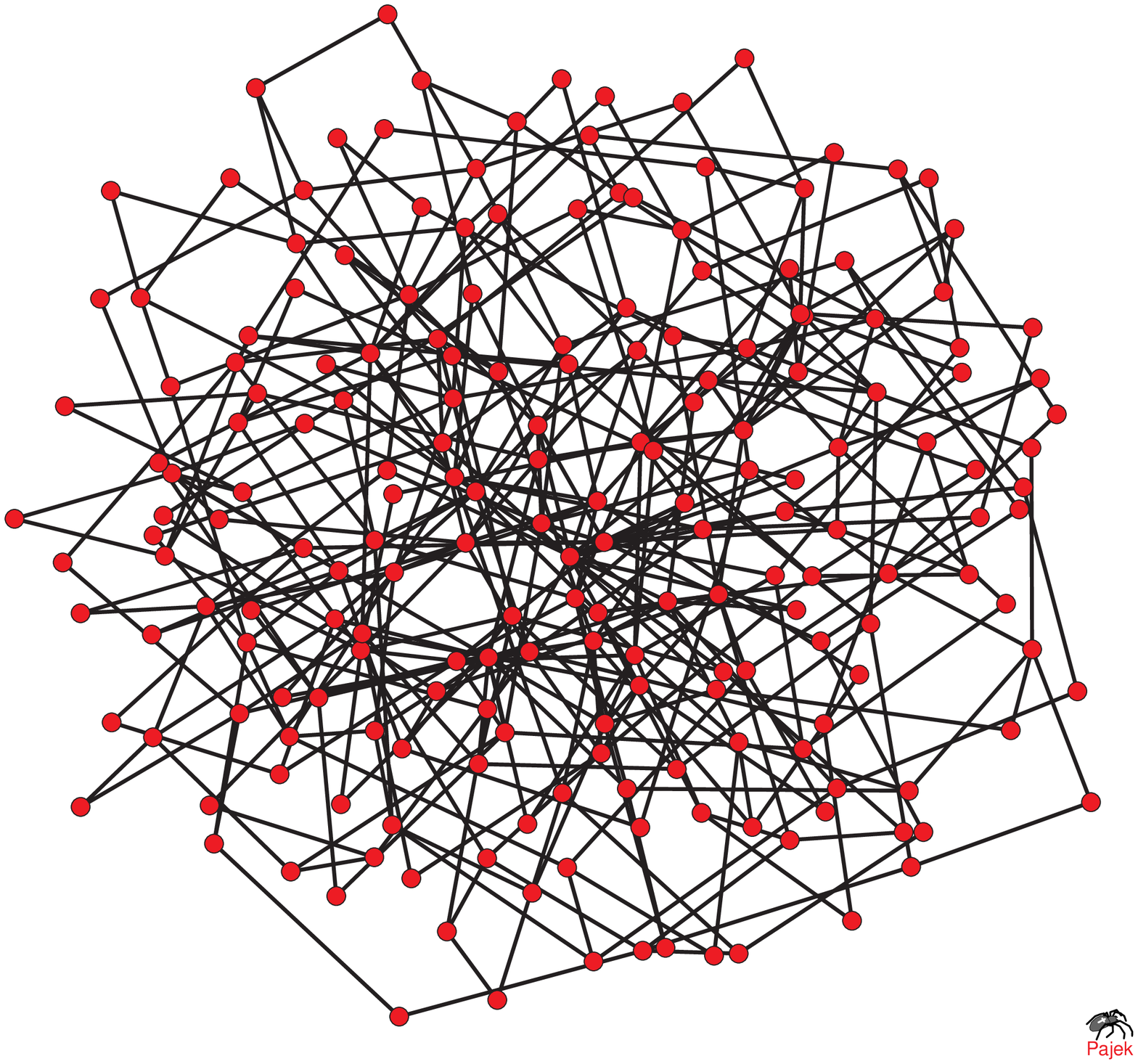}\\
(e) \includegraphics[width=.4\textwidth]{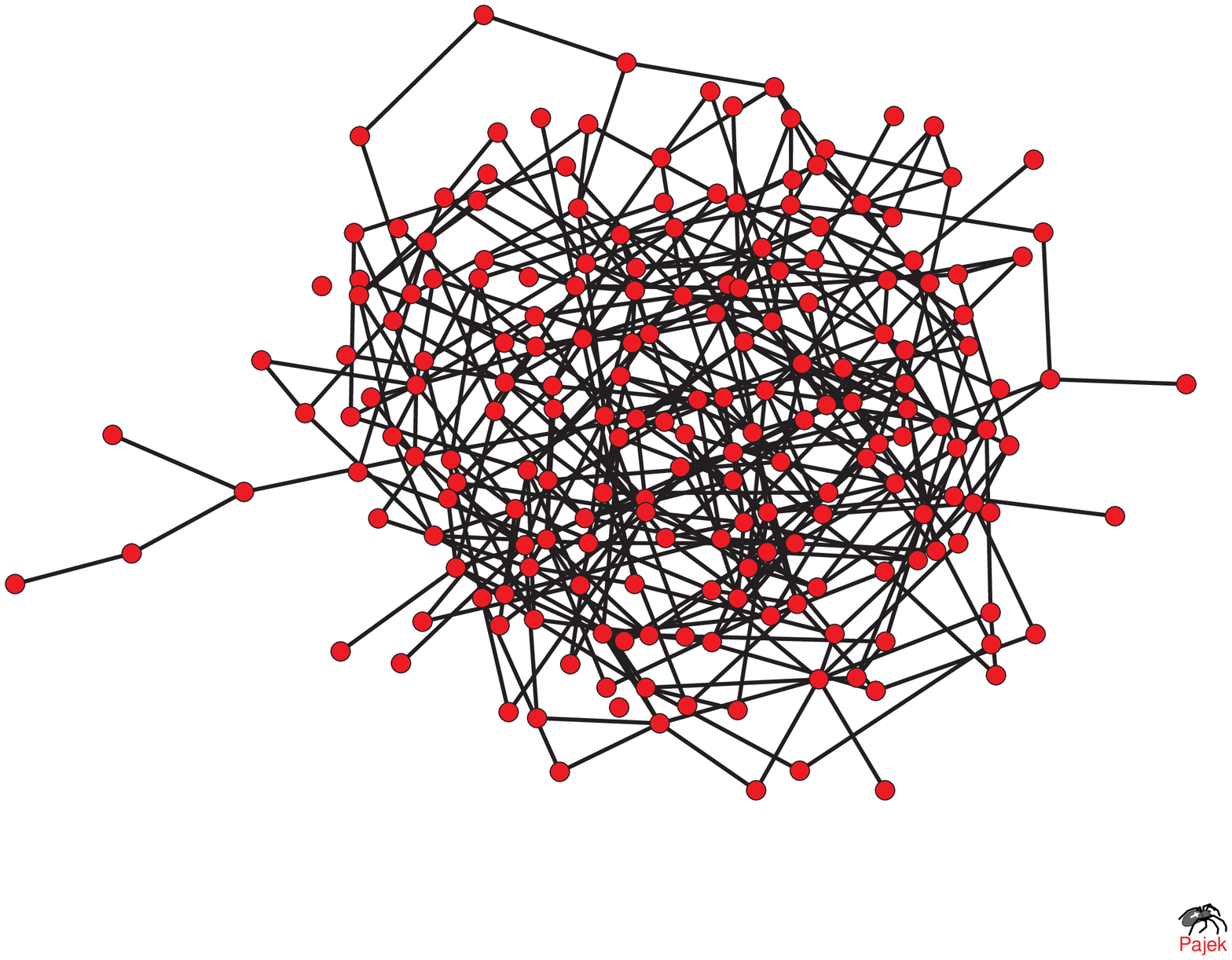}
(f) \includegraphics[width=.4\textwidth]{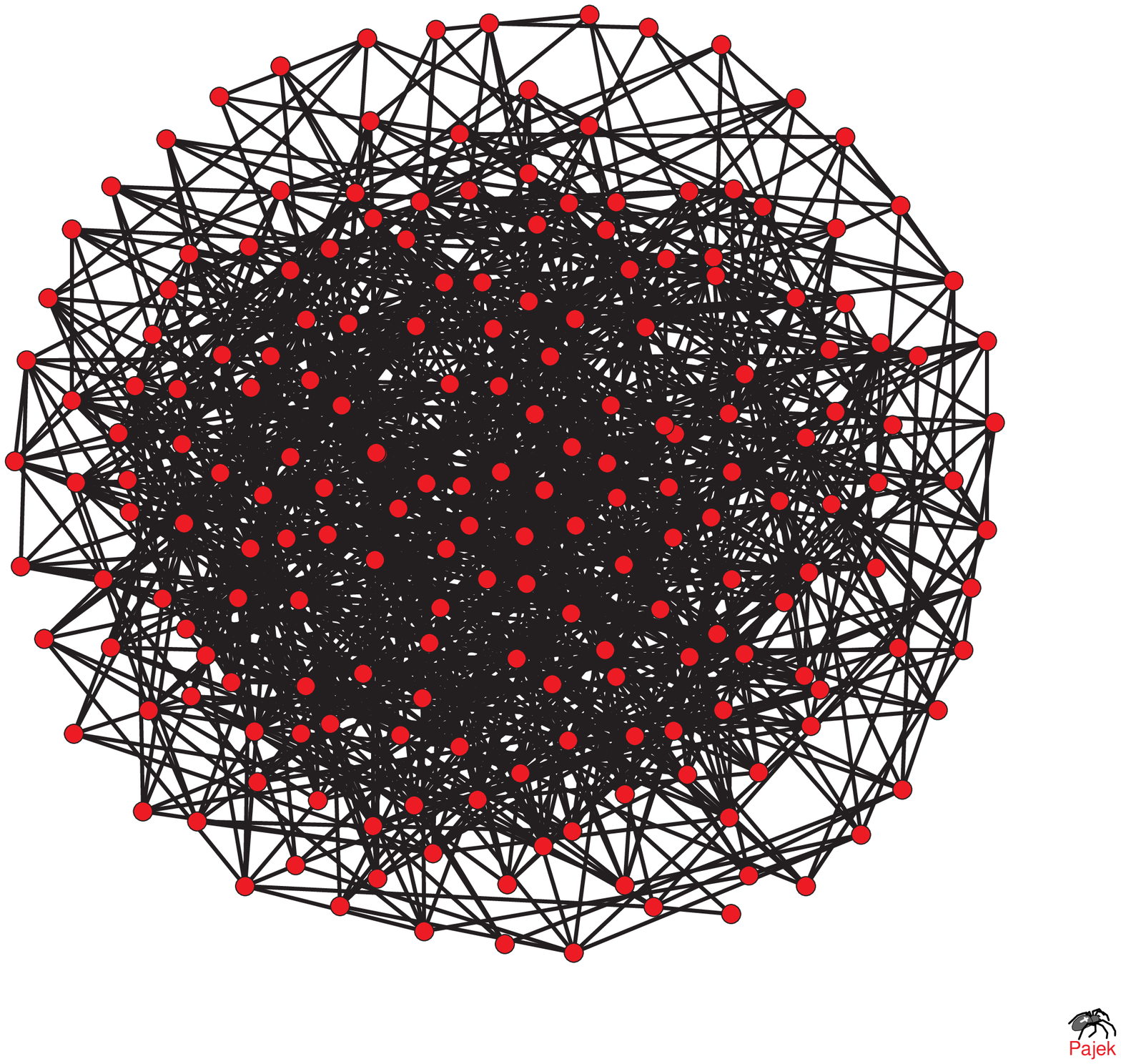}
\end{center}
\caption{\label{fig-examples} Examples of networks for $N=200$ and
(a) scale-free: tree ($M=1$) and
(b) simple graph ($M=2$),
(c) exponential: tree ($M=1$) and
(d) simple graph ($M=2$),
(e) classical random graph: $p=0.02$,
(f) $p=0.05$.
(Figures using Pajek \cite{pajek}.)}
\end{figure*}

In Fig. \ref{fig-examples} examples of scale-free, exponential and CRG networks are presented.

The aim of this work is a brief recapitulation of our recent numerical results, obtained with 
a simple algorithm of network growth, and a comparison of these results with some analytical
predictions. 
The algorithm has been applied to the exponential networks, the scale-free networks and the 
classical random graphs.
The algorithm was verified by a comparison of the results with exact iterative equations.
In Sec. \ref{sec-2} our algorithm is explained.
In Sec. \ref{sec-3} we present the numerical results on the degree and degree distributions
and their correlations, which appear to be relevant for some search algorithms. The distance 
distributions are compared with theoretical formulas.
Final conclusions (Sec. \ref{sec-4}) close the text.

\section{Matrix representation and the growth}
\label{sec-2}
A network containing $N$ nodes is fully characterized by its {\em adjacency matrix} $\mathbf{A}$,
 elements of which give number of edges between nodes $i$ and $j$.
In case of simple graphs --- where multiple edges are forbidden --- this matrix becomes binary: 
$a_N(i,j)=1$ if the nodes $i,j$ are linked together, and $a_N(i,j)=0$ elsewhere.
Absence of loops gives all diagonal elements equal to zero: $a_N(i,i)=0$.

In {\em the distance matrix} $\mathbf{D}$, the matrix element $d_N(i,j)$ is the number of links 
along the shortest path from node $i$ to $j$.

The conversion $\mathbf{D}\to\mathbf{A}$ is trivial, as we need only change all elements larger 
than one to zero.
For building the distance matrix $\mathbf{D}$ basing only on the adjacency matrix $\mathbf{A}$ 
is more complicated and several numerical techniques are available \cite{atod}.
These methods usually base on the list approach with the breadth-first search or depth-first 
search or subsequent usage the Dijkstra algorithm \cite{alg} for all nodes.

Here, we present algorithms for different kind of networks which allow for a construction of 
the distance matrices {\em simultaneously} with the network growth, and not afterward 
\cite{ijmpc,physicaa,task,epjb}.

\subsection{Numerical approach}

For growing networks, the starting point is a matrix $\mathbf{D}$ for the tree of two nodes 
linked together:
\[
\mathbf{D}_2=
\begin{pmatrix}
0&1\\
1&0\\
\end{pmatrix}
.
\]

Selecting a node $m$ to which a new node will be attached is equivalent to select a number $m$ 
of column/row of the matrix $\mathbf{D}$.
Approaches presented here base on the fact that the distance $d(n,i)=d(i,n)$ to a new node $n$
 from all other preexisting nodes $i$ {\bf \em via} node $m$ is $d(m,i)+1=d(i,m)+1$.

\subsubsection{Growing trees}
\begin{figure*}
\begin{center}
\includegraphics[scale=.5]{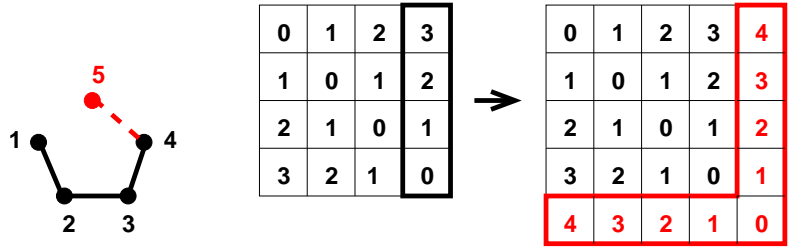}
\end{center}
\caption{\label{fig-D-tre} One step of the distance matrix $\mathbf{D}_4\to\mathbf{D}_5$ 
evolution for growing trees.}
\end{figure*}

Let us start with the simplest case, i.e. when $M=1$.
In this case a tree appears.
Subsequent stages of distance matrix $\mathbf{D}$ evolution $\mathbf{D}_N\to\mathbf{D}_{N+1}$ 
for $N\ge 2$ are
\begin{subequations}
\begin{equation}
\begin{split}
\forall ~ 1\le i\le N: d_{N+1}(N+1,i)=d_{N+1}(i,N+1)=\\
d_N(m,i)+1
\end{split}
\end{equation}
and for diagonal element
\begin{equation}
d_{N+1}(N+1,N+1)=0
\end{equation}
since we do not allow for loops \cite{ijmpc,task}.
\end{subequations}
One step of the distance matrix $\mathbf{D}$ evolution for growing trees is presented in 
Fig. \ref{fig-D-tre}.

\subsubsection{Growing simple graphs}
\begin{figure*}
\begin{center}
\includegraphics[scale=.5]{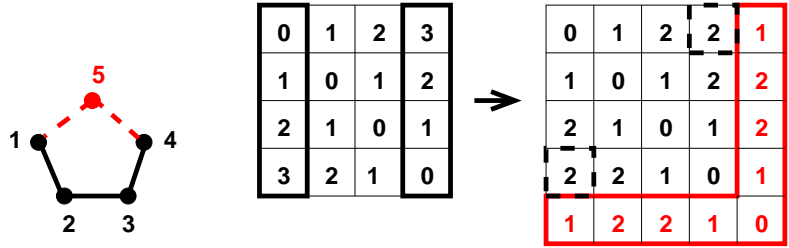}
\end{center}
\caption{\label{fig-D-gra} One step of the distance matrix $\mathbf{D}_4\to\mathbf{D}_5$ 
evolution for growing simple graphs.}
\end{figure*}

In simple graphs cyclic paths are possible and the distance matrix $\mathbf{D}$ is to be 
rebuilt when adding a new node \cite{physicaa,task}.
Suppose that a $(N+1)^{\text{th}}$ node is added to existing nodes $m$ and $n\ne m$.
Then, we have:
\begin{subequations}
\begin{equation}
\begin{split}
\forall ~ 1\le i,j\le N: d_{N+1}(i,j)=\min\big(d_N(i,j),\\
d_N(i,m)+2+d_N(n,j),d_N(i,n)+2+d_N(m,j)\big).
\end{split}
\end{equation}
For new, $(N+1)^{\text{th}}$, column/row
\begin{equation}
\begin{split}
\forall ~ 1\le i\le N: d_{N+1}(N+1,i)=d_{N+1}(i,N+1)=\\
=\min\big(d_N(m,i),d_N(n,i)\big)+1
\end{split}
\end{equation}
and again for the diagonal element
\begin{equation}
d_{N+1}(N+1,N+1)=0.
\end{equation}
\end{subequations}
Example of the distance matrix $\mathbf{D}$ evolution for a simple graph is presented in Fig.
 \ref{fig-D-gra}.

\subsubsection{Classical random graphs}
For CRG the starting point of simulations is  an $N\times N$ matrix with all non-diagonal 
elements equal to $N$.
 Note that $N$ is larger than any possible distance in connected graphs of $N$ nodes.
Now we go through all upper-diagonal elements of $\mathbf{D}$ and set $d(i,j<i)$ equal to 
one with the probability $p$ --- basing on the CRG's definition.
Obviously, the matrix $\mathbf{D}$ is kept symmetric.
Each time, when a new edge is added, we have to rebuild the whole matrix $\mathbf{D}$ due 
to link between nodes $i$ and $j$:
\begin{equation}
\begin{split}
\forall 1\le m,n\le N: d(m,n)=\min \big( d(m,n),\\
d(m,i)+1+d(j,n),d(m,j)+1+d(i,n) \big).
\end{split}
\end{equation}
After this procedure, the matrix $\mathbf{D}_N$ contains elements
equal to $N$ only if the graph is not connected \cite{epjb}.

\subsubsection{The Kert\'esz list}
\begin{figure}
\begin{center}
\includegraphics[scale=.5]{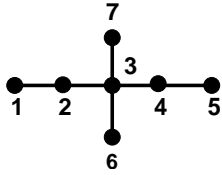}
\end{center}
\caption{\label{fig-kertesz} The Kert\'esz list for small graph presented above 
$\mathbf{r}=\{ 1 ,2 ,2 ,3 ,3 ,4 ,4 ,5 ,3 ,6 ,3 ,7 \}$.
The nodes with lower labels are older.
When a new node is linked to preexisting graph by $M$ edges, $2M$ labels are added to the 
initially existing set $\{1,2\}$.}
\end{figure}

Additional vector $\mathbf{r}$ of nodes' labels may be useful.
There, each node's label appears as an element of $\mathbf{r}$ as many times as it is degree 
of that node (see Fig. \ref{fig-kertesz}).
At each time when new $(N+1)^{\text{th}}$ node is added with $M$ links to nodes labeled as 
$m_1,\cdots,m_M$, these $M$ labels are added to the list, as well as the new label $(N+1)$, 
which is added $M$ times.
By using a randomly chosen element of vector $\mathbf{r}$ as an label, to which new node will
 be attached, we realize the Albert--Barab\'asi rule \cite{ab-org} of preferential growth 
\cite{ds-priva}.

\section{Results of the simulations}
\label{sec-3}
When the evolution process is accomplished, the matrix $\mathbf{D}_N$ may provide much 
information on the graph.
Let us denote $[\cdots ]$, $\{\cdots\}$ and $\langle\cdots\rangle$ the averages over 
$N_{\text{run}}$ different matrices, $N$ nodes of network and $N^2$ matrix elements, 
respectively.  
For example, the average number of elements equal to $d$, denoted as $z_d$, in matrix
 $\mathbf{D}$ gives the average number of $d^{\text{th}}$ neighbors for each node 
($d=1$ --- nearest neighbors, $d=2$ --- next-nearest neighbors, $d=3$ --- next-next-nearest 
neighbors, etc.).
The $i^{\text{th}}$ node degree $k_i=\sum_{j=1}^N a_N(i,j)$ is the number of ``1'' 
in $i^{\text{th}}$ row/column of $\mathbf{D}$ (and $z_1=\{ k \}$).
Average distance to node $i$ from all other nodes 
\[ \xi_i \equiv [N^{-1}\sum_{j=1}^N d_N(i,j)]. \]
The network diameter $\ell$ is the mean length of the shortest path between two vertices
\[ \ell\equiv [N^{-1}\sum_{i=1}^N\xi_i]=[\{ \xi_i \}]=[\langle d_N(i,j)\rangle]. \]

\subsection{Distribution of node degrees}
Three kind of networks presented here derive their names from the distribution of node degrees:
\begin{itemize}
\item for the scale-free networks we reproduce $P_k(k)\propto k^{-\gamma}$ with 
$\gamma\approx 2.72$ ($M=1$) and $\gamma\approx 2.63$ ($M=2$), while the theoretical 
value is 3 and it is independent on $M$ \cite{a-b-rew,ab-org}.
The numerical reduction of $\gamma$ is known to be caused by the finite-size effect.
\item for the exponential trees the node degree distribution is verified to be 
$P_k(k)\propto w^{-k}$ \cite{d-m-rew} where $w=2$ for $M=1$ and $w=3/2$ for $M=2$.
\item the degree distribution for CRG follows the Poisson distribution
$P_k(k)=\exp(-\{k\})\cdot\{k\}^k/k!$, with $\{k\}\approx 20$
and $\{k\}\approx 50$ when $N=10^3$ for $p=0.02$ and $p=0.05$, respectively.
Here, the average node degree may be evaluated as $\{ k \}=p(N-1)$ \cite{crg}.
\end{itemize}

\begin{figure}
\begin{center}
\includegraphics[width=0.45\textwidth]{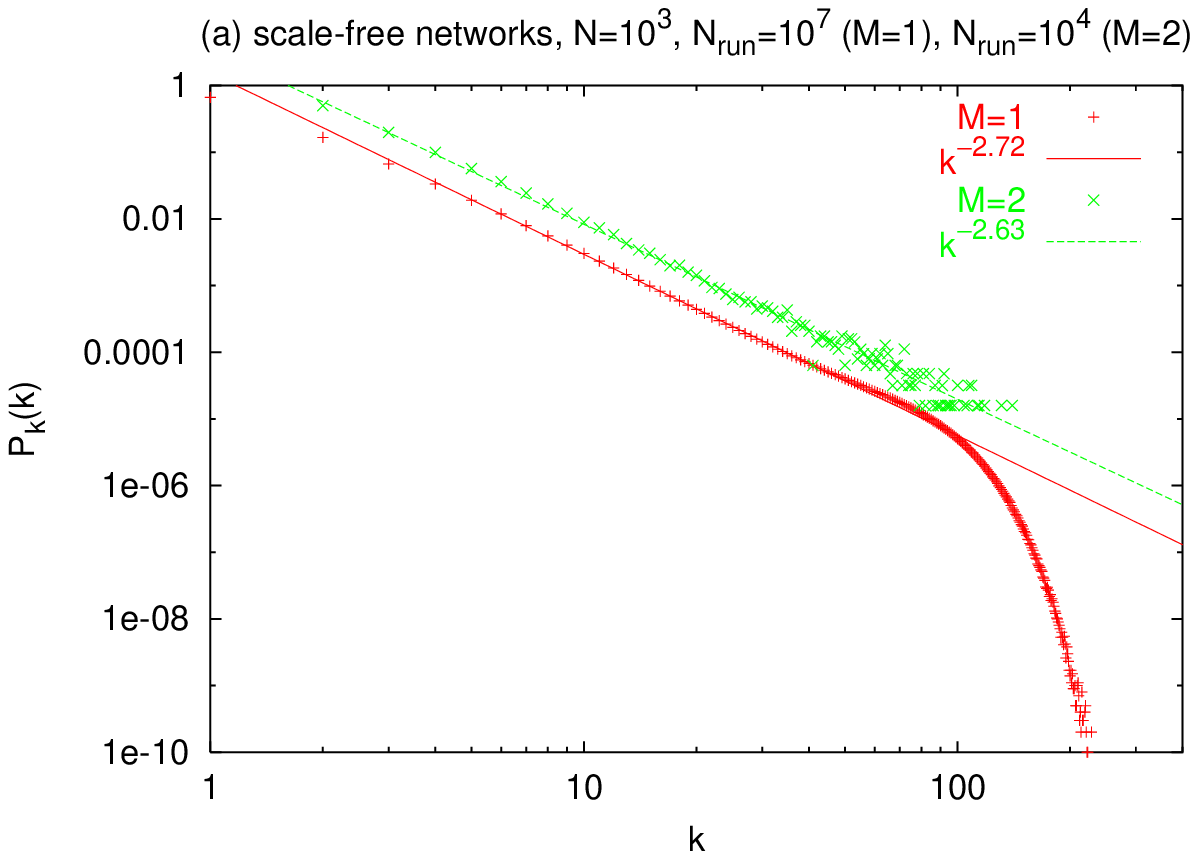}
\includegraphics[width=0.45\textwidth]{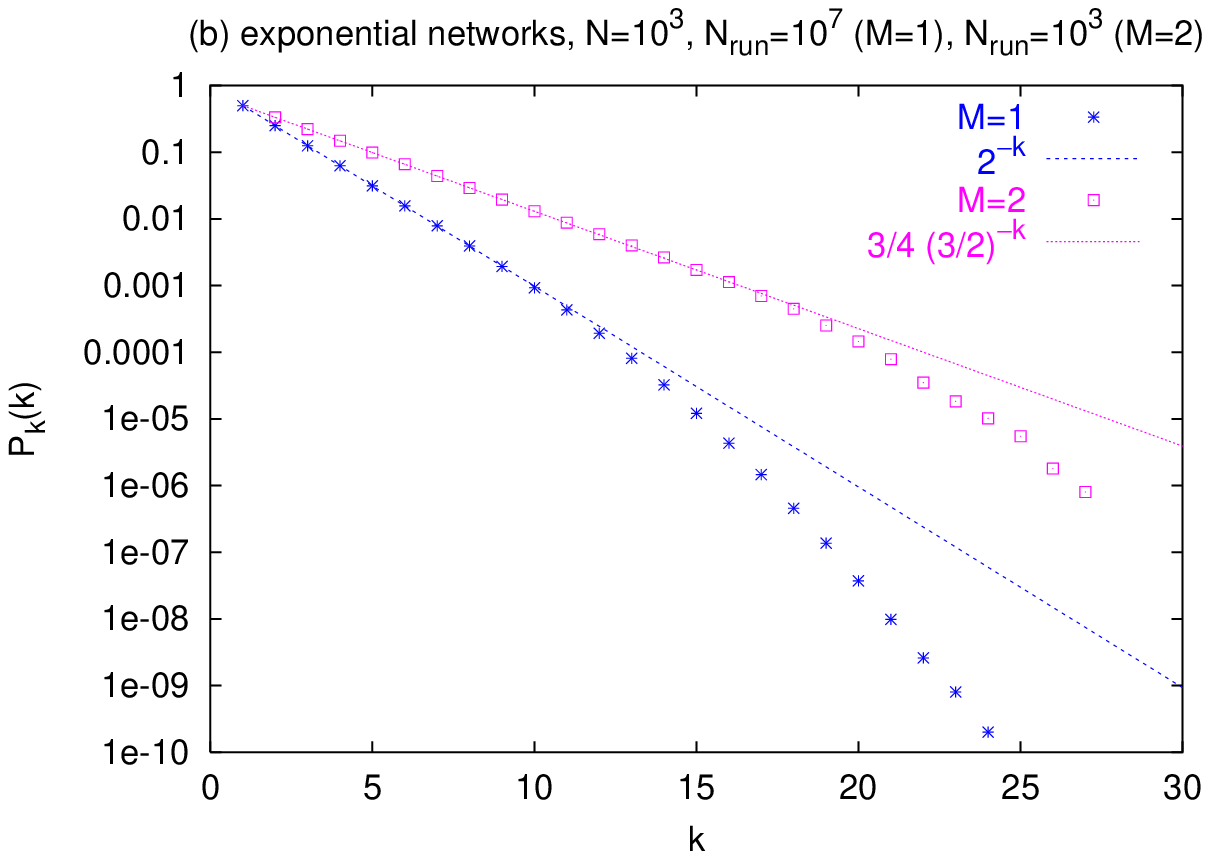}
\includegraphics[width=0.45\textwidth]{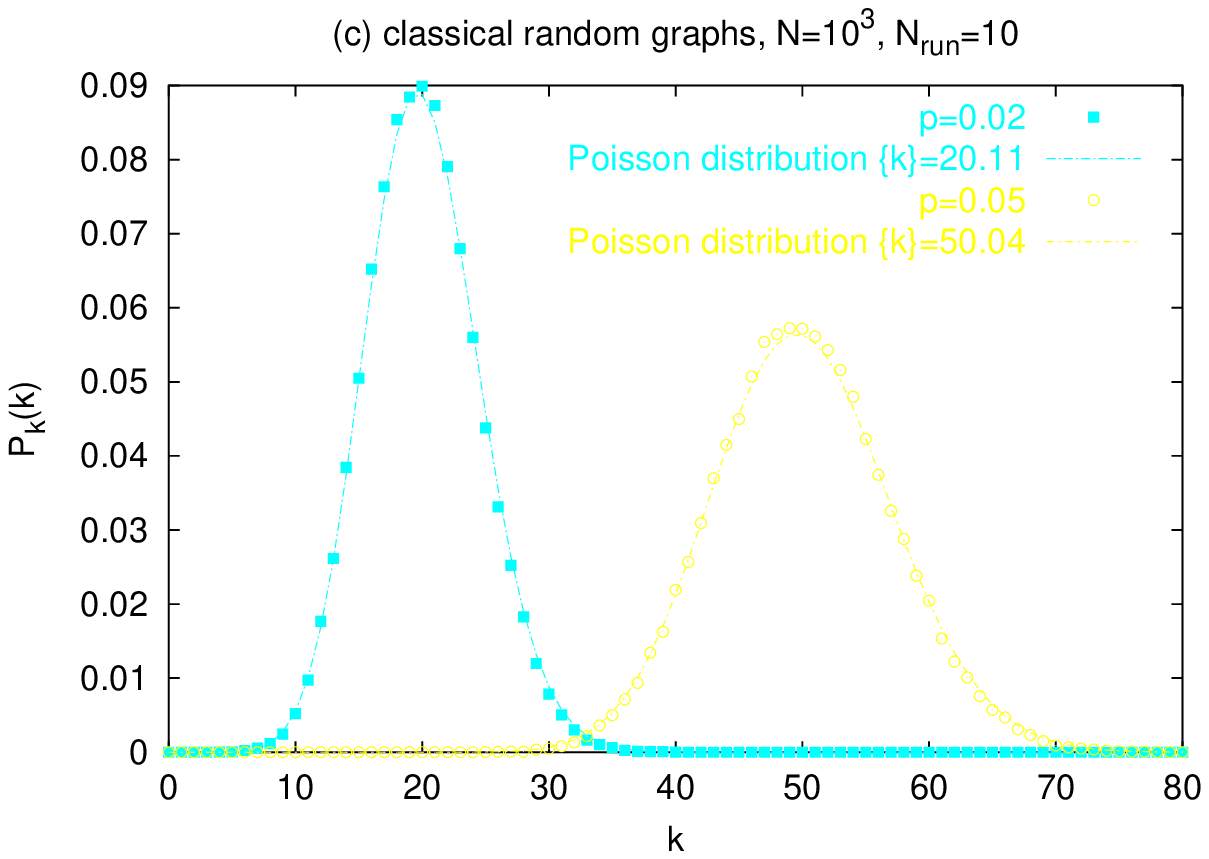}
\caption{Nodes' degree distribution $P_k(k)$ for (a) scale-free, (b) exponential and (c) CRG.}
\label{fig-degrees}
\end{center}
\end{figure}

\subsection{Distribution of node-to-node distances}
\begin{figure}
\begin{center}
\includegraphics[width=0.45\textwidth]{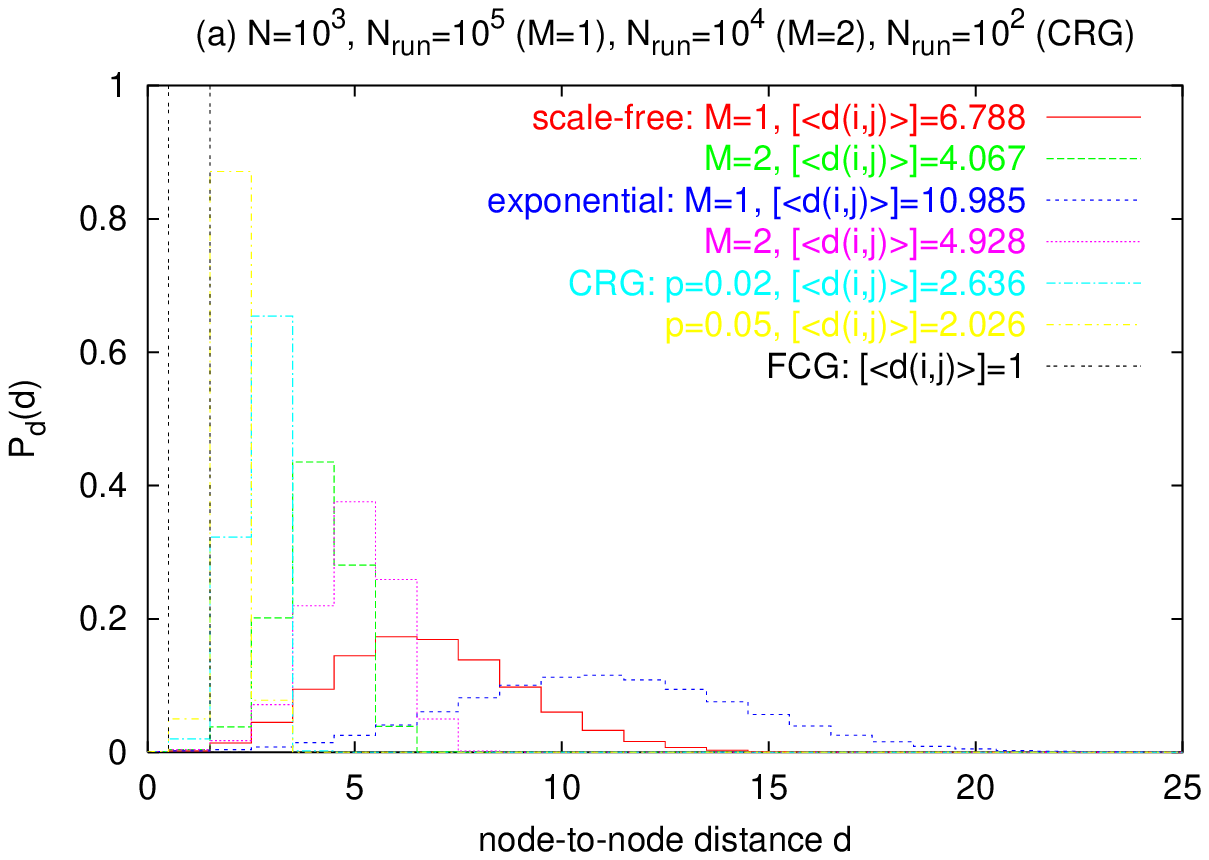}
\includegraphics[width=0.45\textwidth]{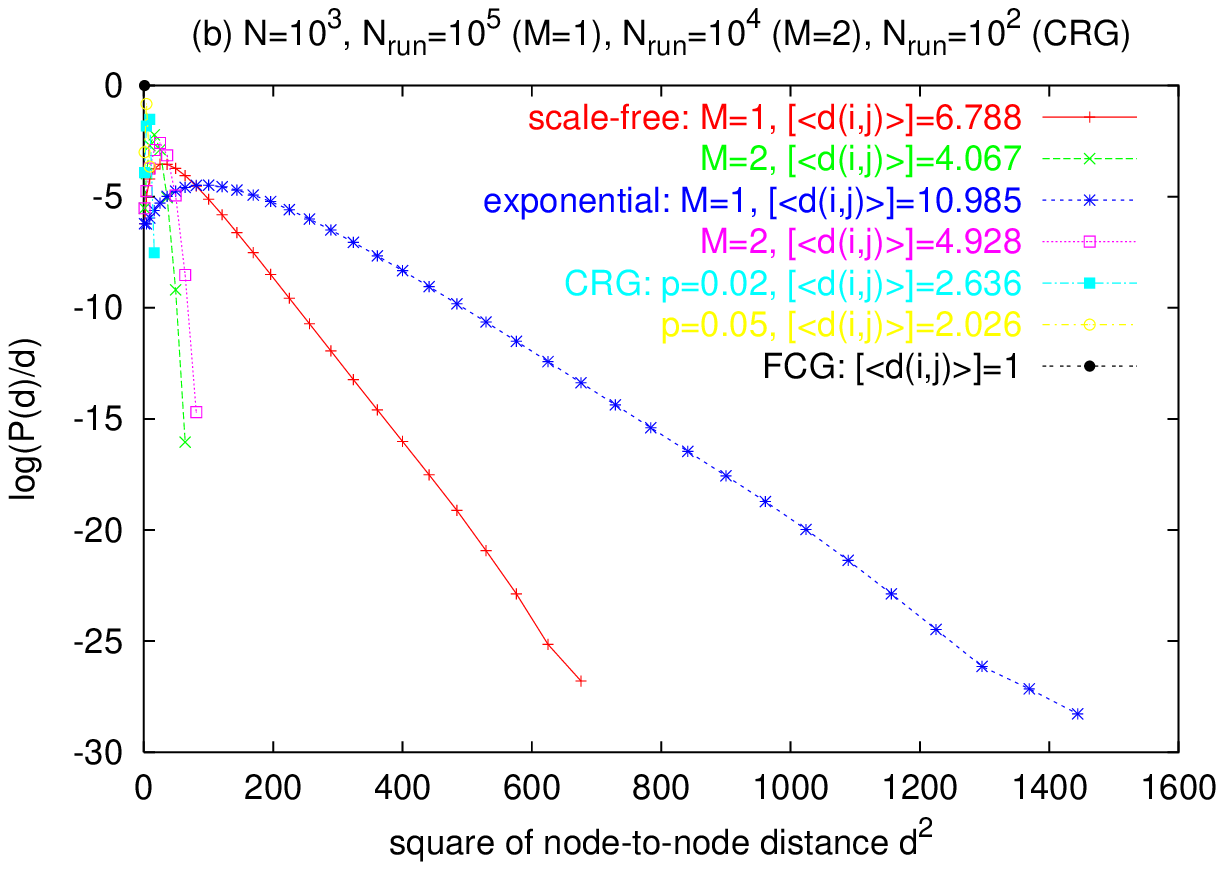}
\caption{Distribution of node-to-node distances.}
\label{fig-distances}
\end{center}
\end{figure}

Distribution of numbers $d$ in distance matrix $\mathbf{D}$ gives node-to-node distance 
distribution (NNDD) $P_d(d)$.
As expected, NNDD for the simple graphs are more condensed than NNDD for trees. Also, NNDD 
for the scale-free graphs (trees) are more condensed than NNDD for the exponential graphs 
(trees) \cite{task} [see Fig. \ref{fig-distances}(a)].
As a rule, networks with more condensed NNDD are themselves more compact.
For large trees, NNDD may be approximated \cite{burda-2} as $P_d(d)\propto d\exp(-Ad^2)$ --- 
see Fig. \ref{fig-distances}(b).

\begin{subequations}
For exponential trees --- where each node is chosen with the same probability as a potential 
candidate to which the next attachment will take place --- it is possible to derive 
\cite{ijmpc,task} exact iterative formula for the network diameter:
\begin{equation}
(N+1)^2\ell(N+1)=N(N+2)\ell(N)+2N
\end{equation}
and for all higher moments $n\ge 2$ of NNDD
\begin{equation}
\begin{split}
(N+1)^2 [\langle d_{N+1}^n(i,j) \rangle]=
N(N+2) [\langle d_N^n(i,j) \rangle]+\\
+2N \sum_{k=1}^{n-1} {n \choose k} [\langle d_N^k(i,j) \rangle]
+2N.
\label{eq-iter}
\end{split}
\end{equation}
\end{subequations}

\begin{figure}
\begin{center}
\includegraphics[width=0.45\textwidth]{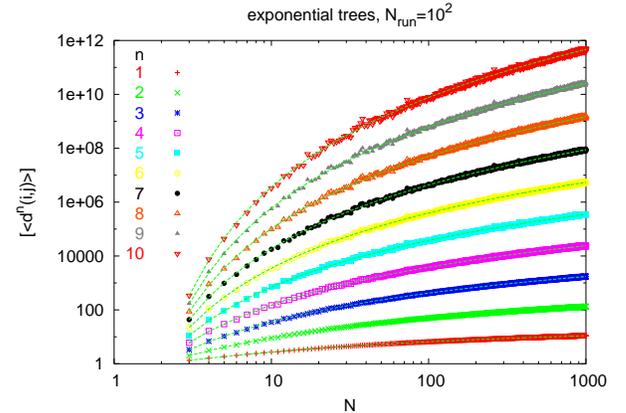}
\caption{Dependence of $n^{\text{th}}$ moment $[\langle d^n(i,j) \rangle]$ for $n=1,\cdots,10$ 
on network size $N$ for the exponential tree.
The lines are given by Eq. \eqref{eq-iter}}
\label{fig-kardas}
\end{center}
\end{figure}

\subsection{Search for a shortest path}

Shortest distance between nodes, averaged over the network is termed the network diameter. 
This diameter appears to be surprisingly short in the growing networks: it increases with the 
number of nodes $N$ as slowly as $\log(N)$. This is known as the small world effect \cite{watts}.
For example, for over 800 million web pages in Internet (in 1999) you need, on average, only 
nineteen clicks to reach any of them starting with randomly selected home page of your browser 
\cite{swe-web}. Table \ref{tab-swe} shows the parameters $\alpha$ and $\beta$ in the logarithmic 
law $\ell(N)=\alpha\ln N+\beta$ for various growing networks \cite{ijmpc,task,physicaa}, 
obtained numerically.

The small world effect has its sociological counterpart, discovered by Milgram \cite{milgram}
in 60's. Several persons had to send letters to a dealer in Boston, unknown to them, using their
acquaintances. It was found that in the average, a chain only six links was sufficient to get the 
target. The essential point in the strategy is the search for the next person. A choice --- almost
obvious --- is to select a person most popular and famous (i.e. the node with the highest degree 
among your nearest neighbors). 

\begin{table}
\caption{\label{tab-swe} The mean distance $\ell(N)=\alpha\ln N+\beta$ for different evolving 
scale-free and exponential networks.}
\begin{ruledtabular}
\begin{tabular}{r cccc} 
           & \multicolumn{2}{c}{scale-free} & \multicolumn{2}{c}{exponential} \\
\hline
$M$        & 1       & 2      & 1       & 2  \\
$\alpha$ & 1.00    & 0.48   & 2.00    & 0.71 \\
$\beta$  & $-0.08$ & 0.83   & $-2.84$ & 0.16 \\ 
\end{tabular}
\end{ruledtabular}
\end{table}

\begin{figure}
\begin{center}
\includegraphics[width=.45\textwidth]{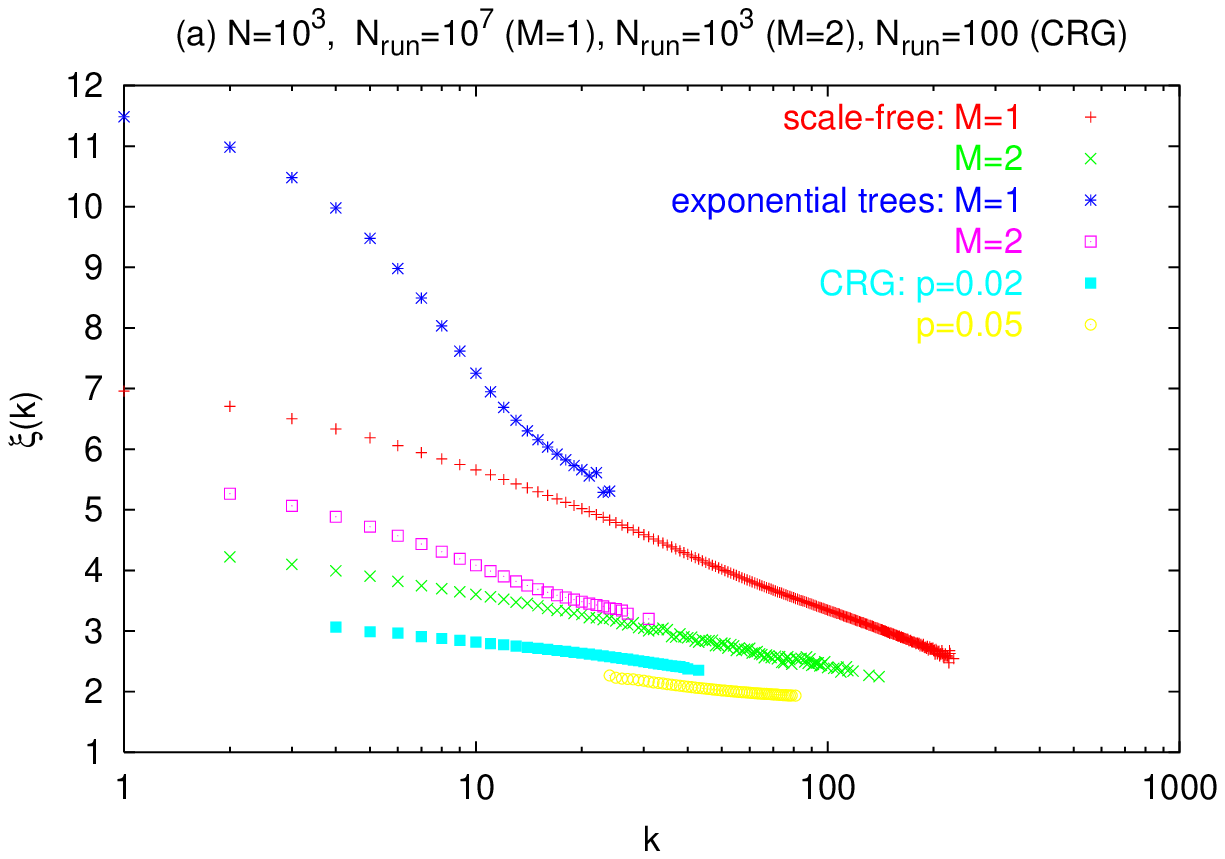}
\includegraphics[width=.45\textwidth]{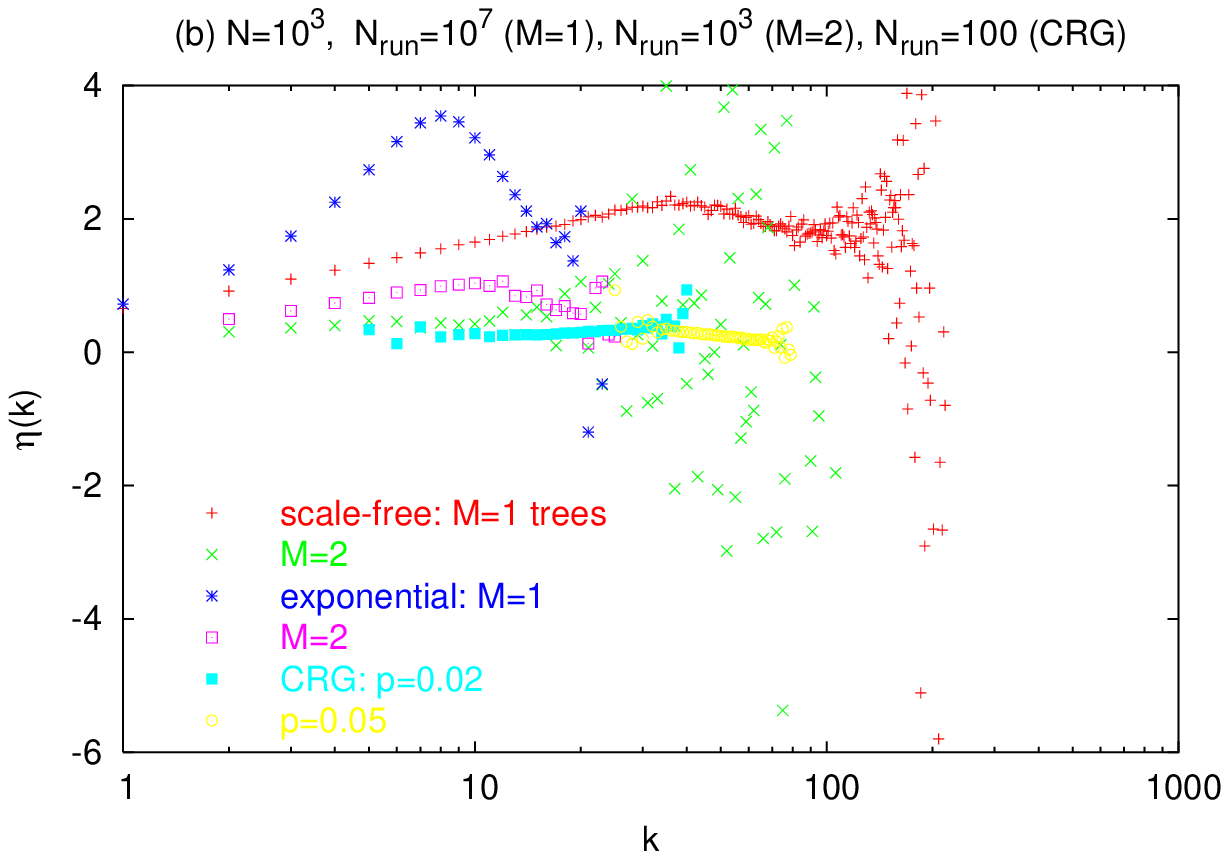}
\caption{Dependence of (a) to-node distance $\xi(k)$ and (b) MCNS efficiency $\eta(k)$ for 
various kinds of networks.}
\label{fig-epjb}
\end{center}
\end{figure}

This strategy is termed as {\em the most connected neighbor search} (MCNS) \cite{mcns}.
The dependence of the average distance $\xi$ from the node on given degree $k$ to all 
other nodes for various networks is presented in Fig. \ref{fig-epjb}(a) \cite{epjb}.
The slope of the curve $\xi(k)$ brings an information, how this search strategy is 
{\em effective} for a given network.
Thus, an effectiveness of MCNS for nodes of given $k$ can be evaluated by an index
\begin{equation}
\eta=-\frac{\partial\xi}{\partial\ln k}.
\end{equation}
The dependence $\eta(k)$ is presented in Fig. \ref{fig-epjb}(b).

The MCNS is more efficient for the exponential trees than for the scale-free trees, and much 
more efficient for trees ($M=1$) than for simple graphs ($M=2$).
For the simple graphs, this search strategy is almost as inefficient as for CRG.
In the scale-free networks, local
fluctuations of degree are enhanced by subsequent linkings.
Multiple centers of high degree can be
created, and the growing concentrates on these centuries.
Then, MCNS can be misleading, as it leads always to a local
center; however, sometimes the target is somewhere else.
This enhancement is
absent in the exponential networks, and that is why MCNS works better there.
We note that this argumentation works particularly well for trees \cite{epjb}.

\section{Summary}
\label{sec-4}

We have described the algorithm of construction the distance matrix $\mathbf{D}$ of the exponential networks, the scale-free networks and the random networks.  The core idea of the algorithm is that it works simultaneously with the network growth.
The information coded in the distance matrix is equivalent to the information of the network structure.
The algorithm's complexity is of order of $N^2$ for trees, and of order of $N^3$ for other networks.
A next step could be to construct a method of comparison of different networks by a comparison of their distance matrices, to check if the networks are topologically equivalent.
However, the order of rows and columns of the distance matrix is set in accordance with the age of nodes, and this information is not preserved in the network topology.
Then, to compare two distance matrices, one should shuffle all possible permutations of nodes.
Such an algorithm is known to be non-polynomial \cite{garey}, and therefore is not useful.

Topological properties of the networks are analyzed by discussing their distance matrices.
These matrices are found to be a convenient tool to investigate the degree distributions, the distance distributions, the small-world effect and some search algorithms in the networks.

We believe that an understanding of topological properties of complex network is the first step to understanding complex behavior occurring among actors and agents occupying the networks nodes 
\cite{action-net}.
It may bring a useful information for modelers of social and economic systems.

\begin{acknowledgments}
We thank J.~Czaplicki, A.~Kardas and J.~Karpi\'nska for their valuable help and Z.~Burda for fruitful discussions.
K.M. acknowledges support from Krak\'ow Branch of the Polish Physical Society (PTF).
Part of calculations was carried out in ACK-CYFRONET-AGH.
The machine time on SGI~2800 is financed by the Polish Ministry of Science and Information Technology under Grant No. KBN/SGI2800/AGH/018/2003.
\end{acknowledgments}


\end{document}